\documentclass[preprint,preprintnumbers,amsmath,amssymb,floatfix,reprint]{revtex4-1}
\usepackage{graphicx}
\usepackage{epsfig}
\usepackage{color}
\begin{document}
\title{Microscale Motion Control Through Ferromagnetic Films}

\author{Andrea Benassi$^{1,*}$, Johannes Schwenk$^1$, Miguel A. Marioni$^1$, Hans J. Hug$^{1,2}$ and Daniele Passerone$^1$}

\affiliation{1-Empa, Swiss Federal Laboratories for Materials Science and Technology, CH-8600 D\"{u}bendorf, Switzerland.}
\affiliation{2-Department of Physics, Universit\"{a}t Basel, CH-4056 Basel, Switzerland.}
\affiliation{* andrea.benassi@empa.ch}

\maketitle

{\bf
Actuation and control of motion in micromechanical systems are technological challenges, since they are accompanied by friction and wear, principal and well-known sources of lifetime reduction. 
In this theoretical work we propose a non-contact motion control technique based on the introduction of a magnetic interaction, i.e., non-contact magnetic friction. The latter is realized by coating two non-touching sliding bodies  with ferromagnetic films. The resulting dynamics is determined by shape, size and ordering of magnetic domains arising in the films below the Curie temperature.  
We demonstrate that the domain behavior can be tailored by acting on handles like ferromagnetic coating preparation, external magnetic fields and the finite distance between the plates. In this way, motion control can be achieved without mechanical contact. Moreover, we discuss how such handles can disclose a variety of sliding regimes.
Finally, we propose how to practically implement the proposed model sliding system.}\\

\section*{Introduction}

Understanding and controlling the frictional properties of two sliding bodies at the micro and nano-scale can have an impact on many technological applications, such as energy conversion and saving, transportation and micro-machining.
In the micro and nanoworld wear and stiction severely limit the device performance and lifetime. 
The traditional techniques to control friction at the macro-scale, such as lubrication and mechanical or chemical manipulation of the sliding surfaces, do not apply straightforwardly  because of the different scaling of physical laws with the system size. Specifically at the nanoscale, peculiar means to control friction would be available, for instance the 
commensurability of the crystal lattices of the sliding nano-objects and the onset of a superlubric state \cite{vanossi}. However,
once the materials have been chosen, it is hard to modify the contact geometry and no efficient friction control can be achieved \cite{dienwiebel,schirmeisen2}. One possibility to 
circumvent this difficulty is to tune the material properties through some external parameter.
For instance, driving a system through a ferroelectric phase transition can give rise to a non-trivial behaviour of friction at the critical point; friction control can be also reached through 
the application of an external electric field \cite{mio}. Using organic ionic liquids as lubricants, the lubrication properties can be again controlled by an electric field \cite{sweeney}, but such  phenomenology is not easily and completely exploitable toward a full friction control.\\
In this work we move instead to the mesoscale realm, 
namely to the study of two micron-sized plates coated with ferromagnetic films. We demonstrate that we can achieve motion control through magnetic friction without mechanical contact, and thus wear, by means of interacting magnetic domains arising at the surface below the Curie temperature. 
Such domains can be easily controlled in size, shape and ordering by means of an external magnetic field, allowing flexibility and continuous tuneability.  
The typical domain sizes, ranging from tens of nm to tens of $\mu$m, makes the domains particularly suitable for integration into micro-mechanical devices.\\ 
Through the simulation of two bodies coated  with thin ferromagnetic films (FFs), as depicted in Figure 1 (a), we show that the interplay of sliding motion and domain dynamics discloses a variety of non-trivial sliding regimes, in some cases involving plastic deformation of the domains. 
We found, for instance, that a stick-slip motion arises when the magnetic domains are ordered into parallel stripes. Its periodicity is dictated by the characteristic stripe width and resembles the mechanical and magnetic friction behavior present at the nano-scale due to atomic periodicity \cite{vanossi,wiesendanger}.
Phenomena such as superlubric transition, the effect of commensurability and contact geometry, and kink motion, can thus be investigated even in our proposed magnetic mesoscale system.
Similar strategies have been recently proposed, making use of model systems such as ion traps \cite{benassiion,mandelli} and colloidal suspensions \cite{bechinger,vanossicol,binder}, however magnetic domains can
provide more freedom and flexibility, their properties being continuously tunable over many orders of magnitude.\\
We will focus here only on FFs with perpendicular anisotropy, i.e., the easy axis of the magnetization is perpendicular to the film surface. This behavior is typical of Co/Pt or Fe/Ni 
multilayers, permalloy and garnet films, to name a few. 
In these FFs, the domains exhibit metastable disordered maze-like patterns but, under the influence of an external magnetic field, the domains can be ordered into parallel 
stripes or bubble lattices \cite{bertotti,hubert}. The domain size and wall thickness can be controlled by the choice of materials and the film thickness \cite{baltz}. The film deposition rate affects the homogeneity of the FFs, promoting the presence of defects and impurities that serve as pinning sites for the domain walls, thus  controlling the domain mobility \cite{pierce}.
In our model, the measurable material properties are condensed into three tunable parameters with a precise physical meaning. The variation of these parameters allows controlling the domain properties and exploring a ``phase diagram'' presenting several possible sliding regimes. Near the tunable material properties, another ``handle'' is clearly the external magnetic field, which will be tuned within realistic values, and a third one is the distance between the plates. \\
Our theoretical investigation anticipates a forthcoming experimental confirmation of the results illustrated here, and their implementation along both the applied and fundamental lines mentioned above. We conclude by discussing
the possible practical realization of the proposed model sliding system and the different ways to measure its sliding and adhesion properties.
\begin{figure*}
\centerline{\includegraphics[height=8.0cm,angle=0]{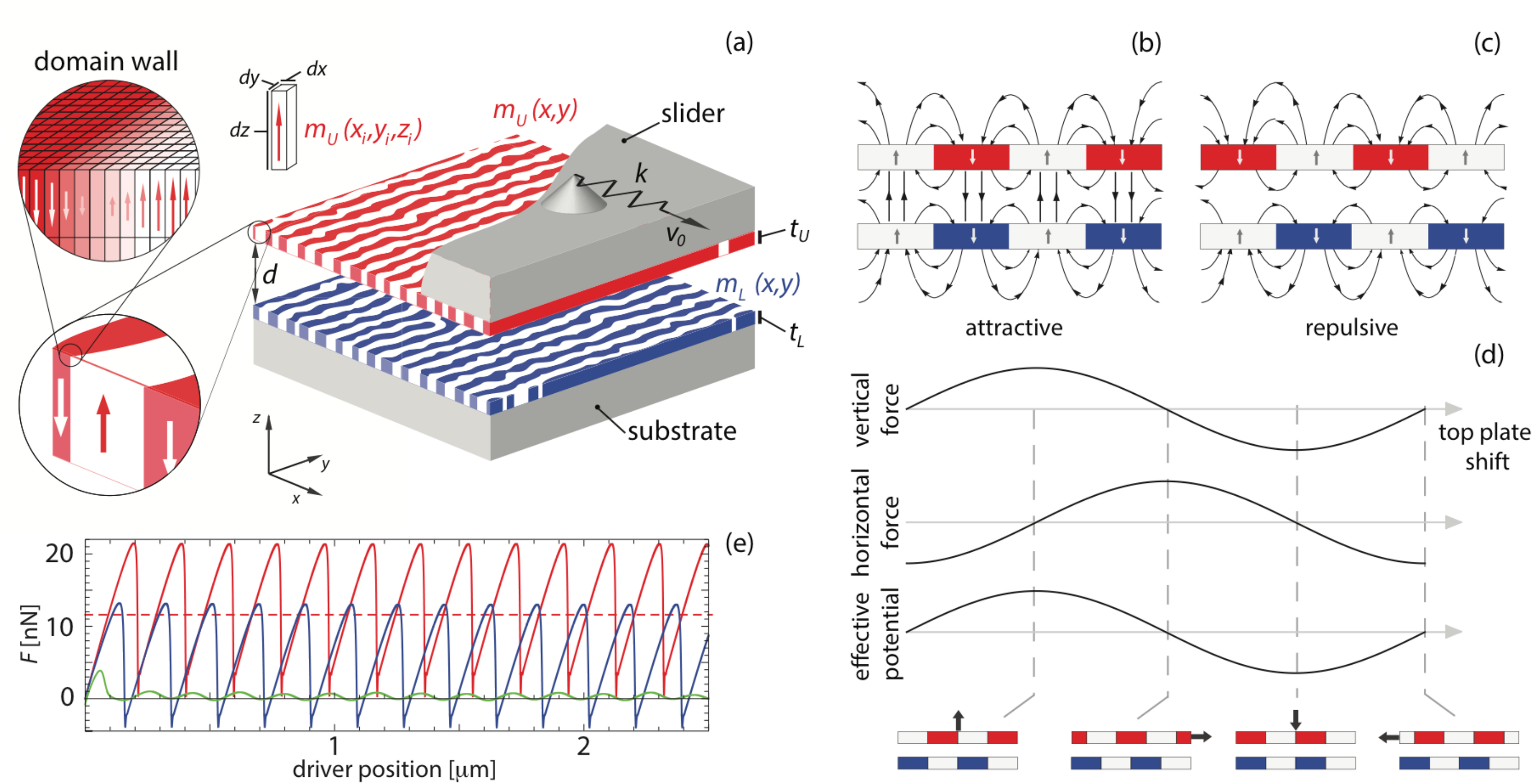}}
\caption{(a) Sketch of the simulated system showing the two supported FFs, their domains and domain walls and the way the magnetization has been discretized in the numerical simulations. (b) and (c) show the minimum 
and maximum energy configurations of the interacting FFs, and their field lines. (d) Force and energy profiles for the interacting FFs.
(e) Force profiles as a function of the driver position (the free end of the spring) for the field values corresponding to the insets of Figure 2 (a), i.e. for stripe domains (red), broken stripes (blue) and bubble lattice (green). The dashed line represents the average
friction force $\langle F(t)\rangle$ or dynamical friction reported in Figure 2.}
\label{figure1}
\end{figure*}

\section*{Results and Discussion}
After a description of the model used for our simulations we introduce the general concepts of magnetic  friction in our system. We then describe in detail the frictional behavior as a 
function of the external field and of the material properties. In a following section we unravel different sliding regimes as a function of the distance between the plates and of the film 
disorder. The latter results are summarized in a ``phase diagram'' (the expression is quoted because of the dependence on the initial and driving conditions) in Figure 3(d), realized by keeping the same external driving conditions and initial domain 
configurations. A final section is devoted to the discussion of adhesion properties.\\

\textbf{Model description.}
For the sake of simplicity and generality we model the geometry of two flat plates sliding parallel to each other at a distance $d$. Magnetic domains are called into play, coating the two bodies with
a lower and an upper FF of thickness $t_L$ and $t_U$ respectively, as sketched in Figure 1 (a). Driving the upper plate laterally we can calculate its resistance to motion due to
the domain interaction, i.e. a \emph{magnetic friction force}. Concurrently we let the domains evolve under the influence of their magnetic fields, varying in time
due to the sliding motion.
The time evolution of the magnetic domains in each supported film is ruled by a Landau--Lifshitz--Gilbert equation (LLGE), describing the precession 
of the magnetization vector $\mathbf{m}$ associated to each infinitesimal volume element of the medium (see insets of Figure 1), around a local magnetic field 
$\mathbf{B}$ due to the rest of the medium and to external sources \cite{gilbert}.
The properties of the upper and lower films (considered identical), and thus the domain behavior, can be characterized by three material parameters entering  the LLGEs (see Computational Method section).
$K_u$ is the macroscopic uniaxial anisotropy constant of the film and represents the energy gained orienting the magnetization perpendicular to the film plane. Due to defects, impurities, and inhomogeneities, the microscopic 
anisotropy can depart significantly from $K_u$.
To take into account this microscopic disorder, which is responsible for the domain pinning,  we introduce spatial fluctuations of the anisotropy with strength $\eta$. 
The last parameter is the exchange stiffness $A$ representing the energy cost of a local variation of the magnetization. 
For many materials, $K_u$ and $A$ have been tabulated \cite{hubert,bertotti} or are fittable from experiments \cite{benassimag2}, whereas $\eta$ depends on the film growth conditions and can be estimated through a statistical analysis of the Barkhausen 
avalanches \cite{benassimag1}. It is a textbook exercise to demonstrate how these parameters determine the behavior of the domains: both the characteristic domain width $\ell$ and the domain wall energy depend on the product  $\sqrt{K_u A}$, the domain wall thickness is set by the ratio 
$\sqrt{A/K_u }$, while $\eta$ establishes the overall domain mobility and the roughness of the domain boundaries \cite{hubert,bertotti}.\\
The position of the lower film is kept constant while the upper one moves as a rigid body, subject to the magnetic interaction with the lower film, and to an external 
driving force. The magnetic force experienced by the sliding film is obtained by integrating over its volume the gradient of the field generated by the lower one (see Computational Method section).
As with the standard simulation of atomic force microscopy (AFM) or surface force apparatus (SFA) experiments, the external driving occurs through a spring $k$ moving at constant velocity $v_0$, representing the stiffness of the driving apparatus, 
and the magnetic sliding friction force is measured through the spring elongation $F(t)=k(v_0 t -x(t))$, i.e. the resistance of the upper film to the external driving.\\
The energy introduced into the sliding system by the driving force and the external magnetic field is dissipated through the excitation of microscopic degrees of freedom in the media, i.e. magnons, eddy 
currents, phonons. These dissipation channels are effectively introduced in our model through viscous damping terms both in the LLGEs, representing the dissipation due to the 
domain motion, and in the Newtonian equation for the motion of the upper film, representing the dissipation through the driving apparatus (see Computational Method section). Depending on the different simulated sliding regimes, the energy dissipation will occur mainly through one channel or the other. \\ 

\textbf{Magnetic friction behavior.}
When a perpendicular anisotropy FF is uniformly magnetized, it is the magnetostatics analogue of a plane capacitor in electrostatics, i.e. the outer field is zero whereas the inner one is uniform, depending only on the film thickness and the saturation magnetization $M_s$.
It is thus clear that, in the absence of magnetic domains, no interaction takes place between the FFs. More interesting is the regime in which the domains of both films are ordered into parallel stripes by means 
of an external field parallel to the surfaces of the FFs. With this domain arrangement, dragging the slider in the direction perpendicular to the stripes, we found a stick-slip regime with the same periodicity as that of the stripes' lateral width.
The key ingredient for the stick-slip motion to take place is the presence of an effective periodic potential between the substrate and the slider, with alternating minima and barriers. 
To understand why this is the case for stripe domains, 
one can consider the domains as microscopic magnets with a north and a south pole. If the two films expose opposite poles to each other, as in Figure 1 (b), they experience the strongest attractive force along 
$z$ and zero force along $x$. A slight displacement of the upper film produces a restoring force, i.e. this is the most stable configuration, corresponding to an energy minimum. But if the films expose the same pole to each other, as
in Figure 1 (c), they feel the strongest repulsion along $z$ and again zero force along $x$. This corresponds to an unstable configuration as, displacing slightly the system, a force along 
$x$ will drive the system away from its original configuration. The calculated force and potential profiles are sketched in Figure 1 (d), and some examples of a sawtooth friction force signal due to the stick-slip phenomenon are given in Figure 1 (e).\\ 
The domain dynamics is studied with the following simulation protocol. We suppose that the domains have been initially aligned into stripes, we start sliding the upper plate perpendicularly to the stripes direction, and we 
record the magnetic friction force over time. After an initial transient, in which the domains readjust, a steady state is reached and the average magnetic friction $\langle F(t)\rangle$ can 
be calculated. \\ 

\textbf{Controlling magnetic friction with an external field.}
In this section the effect of an external magnetic field on the domain morphology and dynamics, and ultimately on the frictional response of the system, is discussed. The simulation protocol described at the end of the previous section can be repeated for any prescribed value of an external magnetic field perpendicular to the FFs.
\begin{figure*}
\centerline{\includegraphics[height=13.0cm,angle=0]{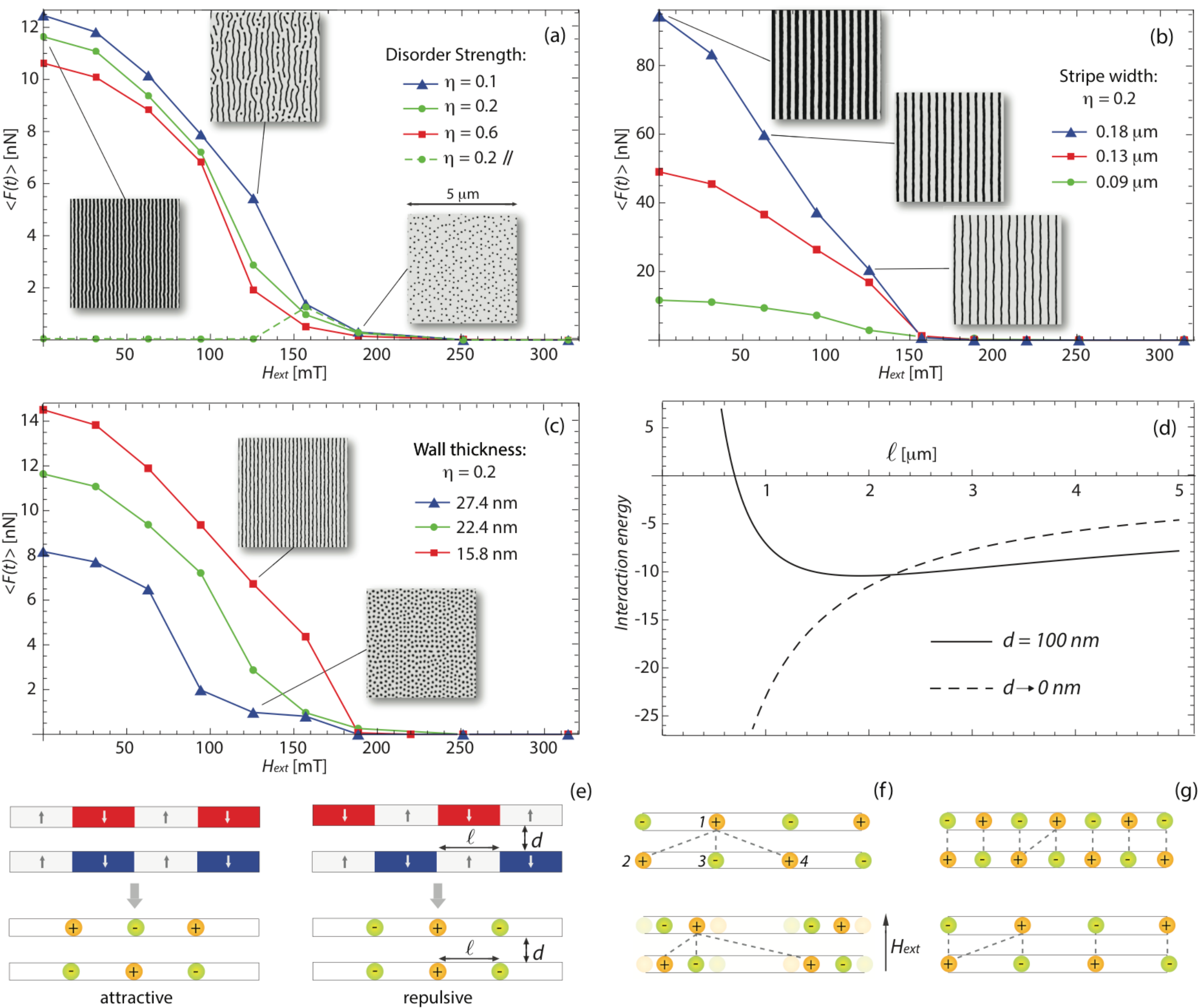}}
\caption{(a) Friction force versus external field as a function of film disorder. (b) Friction force versus external field as a function of stripe domain width. (c) Friction force versus external field as a function of domain wall thickness. 
All the insets are $5\times 5\,\mu$m$^2$ images of the lower film magnetization $m_L(x,y)$ in the steady state while the upper film is sliding.
(d) Interaction energy as a function of the stripe width $\ell$ calculated with the effective model. The energy unit is $\mu_0 M_s^2 t_U t_D L$ with $L$ the total length of the film. 
(e) Sketch of the effective model where domain walls are represented by interaction centers. (f) Effect of an external field on the position of the interaction centers. (g) Effect of the stripe width on the film--film interaction.}
\label{figure2}
\end{figure*}
The average magnetic friction force as a function of the perpendicular external field is 
plotted in Figure 2 for different 
material parameters, retaining the same geometry ($t_U=t_L=50$ nm, $d=100$ nm and film size $5\times 5$ $\,\mu$m$^2$).
The general trend is the same for all plots and is illustrated by the insets of Figure 2 (a): at zero or low field the stripe domains give rise to a stick-slip sliding; 
increasing the field the black stripes shrink until their size becomes comparable to the wall thickness. Here they break down leading to the coexistence of stripes and bubbles and the 
stick-slip is suppressed with a significant reduction of friction; at higher fields only disordered
bubble lattices exist, with an extremely low and disordered friction signal; friction goes to zero when the film is completely saturated.
The simulated friction force profiles as a function of the field are plotted in Figure 1 (e).\\
From Figure 2 it can be noticed that the friction force starts to decrease well before the stripe breakdown.
The latter observation can be easily explained using a simplified model and comparing the energetics of different equilibrium configurations.
We can drop the traditional picture of domains interacting like small classical magnets in favour of an equivalent one considering the total magnetic energy 
as built up by interactions between domain walls. Assuming for simplicity walls of infinitesimal thickness an analytic expression for such contributions can be obtained (see 
Supporting Information).
The wall--wall interaction is attractive if the walls are of the same kind, i.e. both up-to-down or down-to-up, or repulsive if the walls are of opposite kind. For this reason we can replace the walls with 
interaction centers with sign, as in Figure 2 (e) where the minimum energy configuration is represented letting the two films face opposite centers, and the unstable one is obtained facing 
centers of the same sign. 
Without loss of generality we limit ourselves to second-neighbour interactions. In the stable configuration of Figure 2 (f), every upper film interaction center experiences an attractive
interaction with the closest lower film center (1--3) and a repulsive interaction with the two second neighbors (1--2 and 1--4). If we now switch on $H_{ext}$ the stripes shrink or expand,
the 1--3 interaction remains unchanged while 1--2 gets stronger and 1--4 gets weaker. It can be demonstrated that, even for infinitesimal displacements of the center 1, the total repulsive interaction increases provided that $\ell$ 
is of the same order or smaller than $d$ (see Supporting Information for analytical calculations).
Thus, in the presence of an external field, the system gets less stable and offers less resistance to the external lateral force which drives it out of equilibrium, and this results in a decrease of the friction force.\\ 
The frictional behavior as a function of the external field is significantly affected by the properties of the materials. 
In Figure 2 (a) FFs of different film homogeneity are analysed, using the values of $M_s=5\times 10^5$ A/m, $K_u=3.1\times 10^6$ J/m$^3$ and 
$A=2.47 \times 10^{-10}$ J/m typical of Co/Pt multilayers \cite{benassimag2}. For highly homogeneous FFs we have smooth parallel stripes, whereas 
in highly disordered and inhomogeneous FFs the stripes are quite irregular with wiggling boundaries, however the force does not vary significantly. This result clearly shows the robustness of the stick-slip mechanism, which
is not only present in idealized, perfectly geometric domains but also in realistically irregular patterns.
The dashed line in Figure 2 (a) represents the friction force measured sliding parallel to the stripe direction: the friction is zero because of the translational invariance of the interaction energy along the sliding direction.
A non-monotonic behavior of friction is found only at the stripe breakdown, when disordered bubble lattices appear. In this regime, indeed, sliding parallel or perpendicular to the stripes leads to the same friction force.\\
In Figure 2 (b) the role of the stripe width $\ell$ is presented. We varied $K_u$ and $A$ in such a way as to increase the stripe width while maintaining the same wall thickness.
Doubling the width from $0.09$ to $0.18\,\mu$m, the force can increase by almost one order of magnitude. 
Again this behavior can be understood by means of the effective model. 
From Figure 2 (g) we immediately see that enlarging the stripe size, the total repulsive interaction can only decrease, however this also leaves less domain walls on the films, thus reducing the total attractive 
force (see Supporting Information for analytical calculations). Such competing effects give rise to a non-monotonic behavior of the interaction energy, as illustrated in Figure 2 (d). With our choice of $d$ and $\ell$, we lie in the decreasing part of the continuous line, i.e. 
when enlarging $\ell$, the system is more energetically stable offering more resistance to the external force, thus increasing the friction.\\ 
Figure 2 (c) shows the friction behavior as a function of the domain wall thickness upon appropriate variation of $K_u$ and $A$ in order to maintain the same stripe width while varying the wall thickness. As we increase the latter we observe a decrease in the friction force.
This behavior can be understood remembering that the friction force is proportional to the derivative of the magnetic field, which in turn follows the profile of the domain wall: the sharper the domain wall, the more abrupt is the field variation, thus leading to a higher friction force (see Supporting Information for analytical calculations).  
Notice also that the larger the walls, the smaller the minimum external field needed to break the stripe domains, as shown by the insets of Figure 2 (c). In fact, the stripes collapse when their width is comparable with the wall 
size, so narrow walls allow the stripes to persist at larger external fields keeping the friction force larger.\\
All the phenomenology so far described holds also in the case of incommensurate configurations, i.e. when the size and number of stripes in the upper and lower film differ (see Supporting Information). This condition can be easily realized experimentally by taking $t_U \neq t_L$ or by 
using two different materials for the upper and lower films.\\
Movie 1 in the Supporting Information offers an example of how the motion of the upper plate can be dynamically controlled by exploiting the magnetic friction. The sliding starts at $H_{ext}=0$ with stripe domains on both the plates, but during the sliding $H_{ext}$ is suddenly turned on and the 
stripes break down, suppressing the friction and accelerating the motion of the upper plate.\\

\textbf{Film separation and domain plasticity.}
\begin{figure*}
\centerline{\includegraphics[height=10.0cm,angle=0]{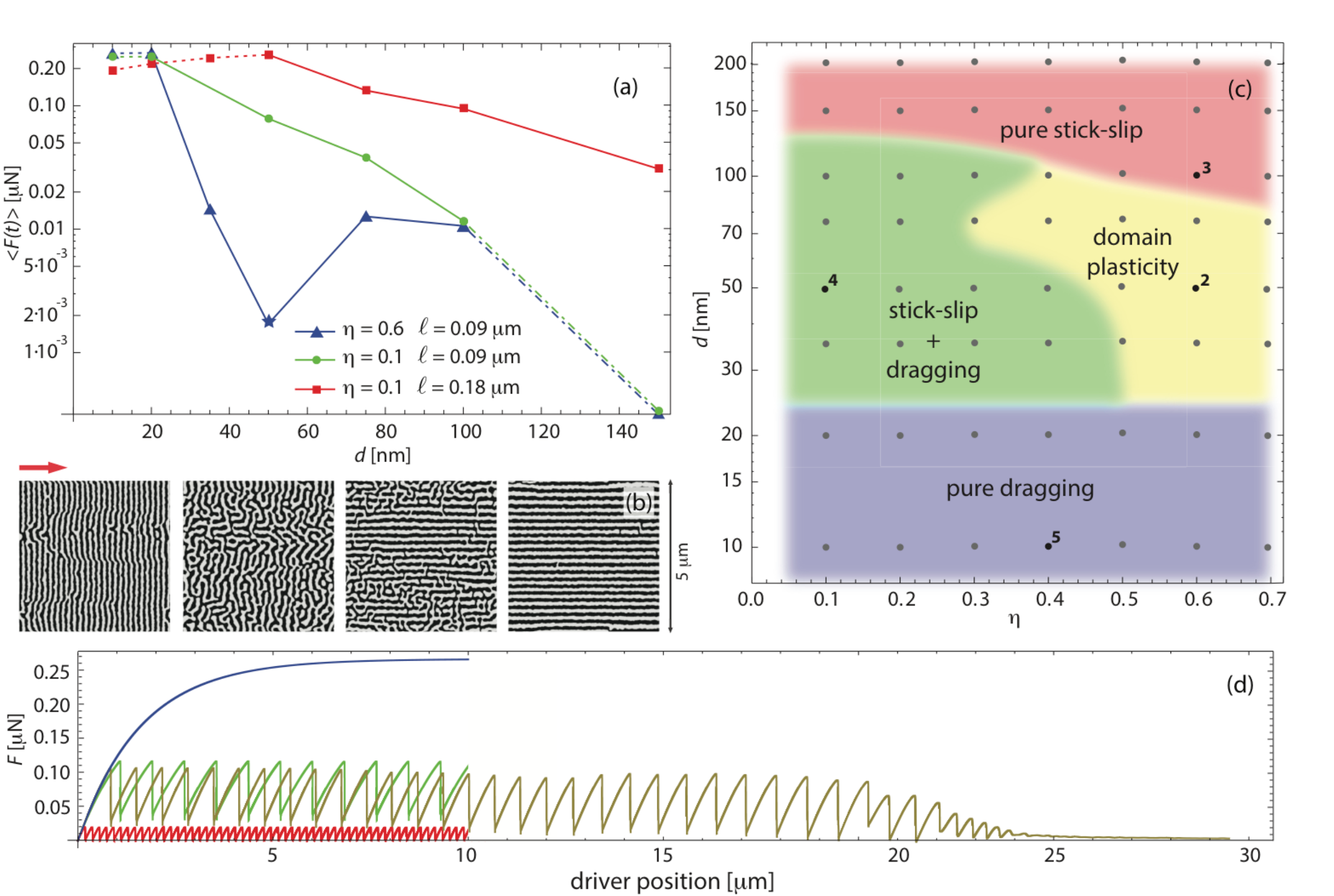}}
\caption{(a) Friction force versus FF separation. (b) Subsequent images of the lower film magnetization $m_L(x,y)$ during the domain rearrangement leading to the force value labelled with a star in panel (a), the arrow indicates the direction of 
The full  $m_L(x,y)$  dynamics is reported in movie 2 in Supporting Information. (c) Sketch of the general domain behavior as a function of the disorder and distance. The domain behavior has been analyzed after the onset of the steady state, always starting from the initial stripe 
pattern with $\ell=0.09 \,\mu$m. The different sliding regimes can also be seen in movies 2-5 in Supporting Information, and the numbers on the plot allow identifying the $d$ and $\eta$ values of each movie.
(d) Friction force profiles corresponding to the points 2-5 of panel (c), the color code is the same. The red curve is the same in Figure 1 (e).}
\label{figure3}
\end{figure*}
The separation $d$ between the FFs can play a significant role in determining the characteristic domain width $\ell$ \cite{baltz}. However, having already discussed how friction depends on the 
latter, we put ourselves in a parameter range where $\ell$ is not significantly affected by $d$ (see Supporting Information).
Unravelling the behavior of sliding friction as a function of the FF separation $d$ leads to the discovery of interesting non-trivial regimes where the mechanical sliding triggers 
adjustments in the domain morphology. 
Inasmuch as the stripe domain pattern of an isolated FF resembles a sinusoidal variation of the magnetization of wave length $\ell$, the FF stray field will decay exponentially as we depart from its surfaces, the decay length being proportional to $\ell$.
This is visible for large $d$ in the logarithmic plot of Figure 3 (a) where the circle and triangle curves, with $\ell=0.09 \,\mu$m, decay more steeply than the square curve, with $\ell=0.18\, \mu$m. Nevertheless significant 
deviations from the exponential decay are visible
for smaller $d$, where the curves depart from the initial straight line trend until they reach a plateau for $d< 20$\ nm. 
The dot-dashed lines represent a pure stick-slip motion whereas, decreasing $d$, the FF interaction increases and the energy barrier to the first slip event grows, as well as the force to be applied. When this force is bigger than the average pinning force that keeps the domains anchored to the FF defects and 
inhomogeneities, the lower film stripes start to move before the slip event take place. This domain 
dragging is however very slow, the magnetization precession being viscously damped (see Computational Method section), so if the driving velocity $v_0$ is larger than the dragging velocity $v_d$, the 
spring of the driving apparatus will keep stretching, increasing the applied force and promoting a slip event.
We have thus a regime of stick-slip friction with a small domain dragging in the lower film, as highlighted in Figure 3 (a) by continuous lines. As we move to smaller $d$ 
the force needed to initiate the slip gets even bigger, increasing the domain dragging velocity. 
When $v_0=v_d$ the driving spring remains uniformly stretched and the stick-slip motion disappears, a sliding regime that is highlighted in Figure 3 (a) by dotted lines.
When the FF inhomogeneity $\eta$ is increased the dragged stripes can encounter strong pinning sites, becoming stretched and then breaking, and this plasticity regime is responsible for the friction 
drop of the triangle curve in Figure 3 (a). Figure 3 (b) shows subsequent images of
the lower film magnetization during the initial moments of sliding: the domain reorganization, occurring in both FFs, ends with the creation of new stripes along the sliding 
direction, in the state of smallest friction force. 
This behavior is the opposite to what typically happens in nano-scale tribological systems where the energy minimization tendency drives the systems to a state of larger friction, as in 
the case of two randomly oriented lattices 
that, during sliding, tilt themselves in such a way as to achieve the maximum of commensurability \cite{dienwiebel}.\\
The friction force profiles corresponding to the stick-slip, stick-slip plus dragging, pure dragging, and plasticity regimes are plotted in Figure 3 (d), while Figure 3 (c) is a phase diagram of the different regimes just described.
Analysing the onset of \emph{stick-sip + dragging} motion, one can get information about the average pinning strength, something hardly accessible by direct mechanical measurements until now.
From the onset of \emph{pure dragging} one can infer information about the characteristic damping of the magnetization motion, i.e. about the 
characteristic time for the energy dissipation of the microscopic decrees of freedom, something typically measured with resonance experiments. Finally, from the onset of \emph{domain plasticity} we can 
gain information about the domain effective elasticity and the capability of domains to bear stress.\\

\textbf{Controlling adhesion.}
\begin{figure*}
\centerline{\includegraphics[height=7.5cm,angle=0]{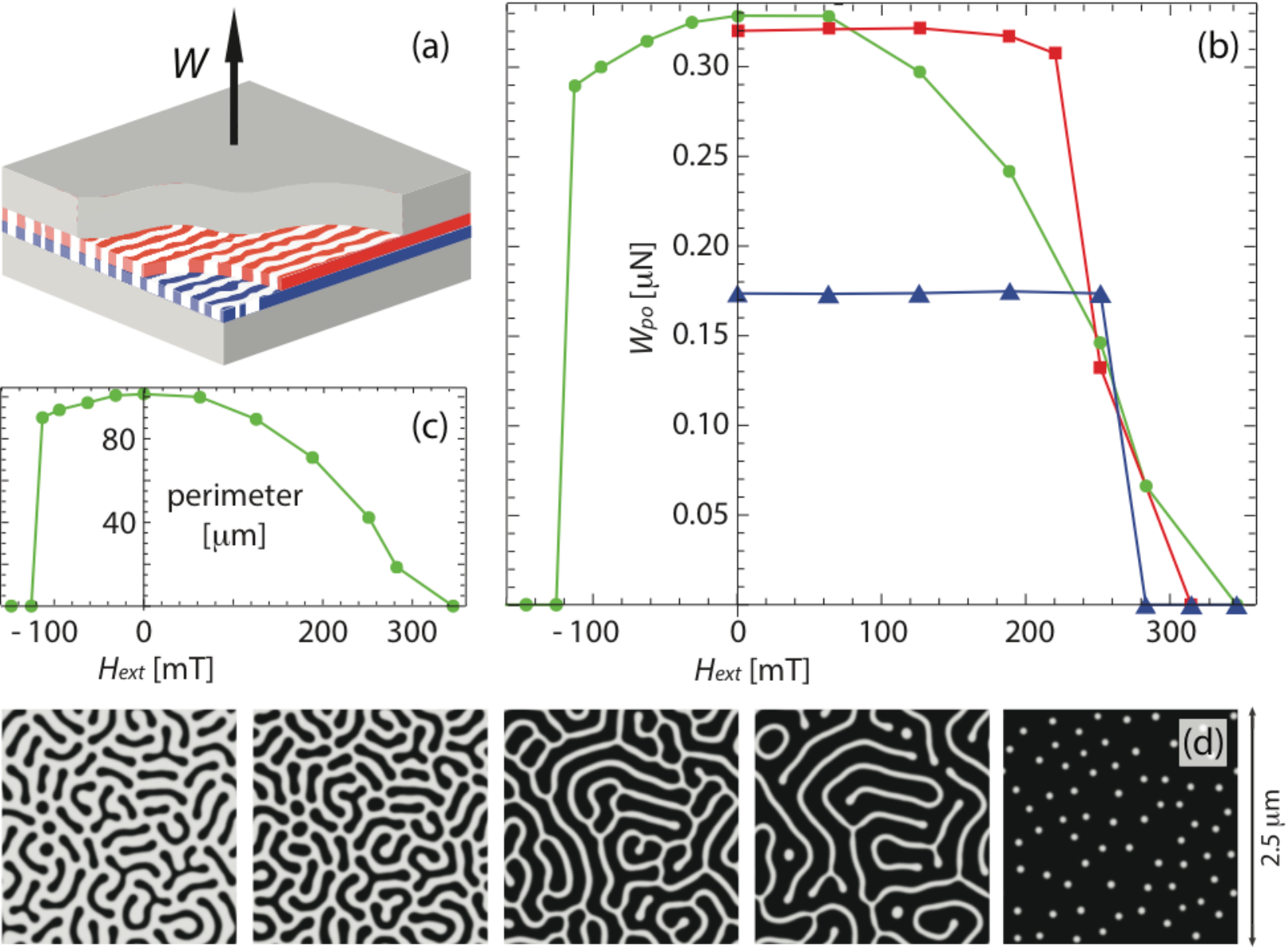}}
\caption{ (a) Sketch of the simulated pull-off experiment. (b) Pull-off force as a function of the external field, starting from $0.09\,\mu$m stripe domains at zero field (squares) and starting from $0.18\,\mu$m stripe domains at zero field 
(triangles), starting from saturated FFs at negative field (circles) with the same parameters as the square curve. (c) Domain perimeter as a function of the applied field for the circles curve of panel (b).
(d) $2.5\times 2.5\,\mu$m$^2$ images of the lower film magnetization $m_L(x,y)$ at equilibrium, prior to the force application. $H_{ext}$ increases from left to right starting from slightly negative values up to the positive saturation.}
\label{figure4}
\end{figure*}
In the limit of zero distance, another class of phenomena can be examined, this time using a vertical instead of a lateral external force. 
Adhesion happens when placing the two films in close contact and ramping up quasi-statically a vertical force $W$ until, at a threshold value $W_{po}$, pull-off takes place, as depicted in Figure 4 (a). 
At $d \rightarrow 0$ each film tends to force its magnetization pattern on the other one, and even in the case of two very different FFs, the same domain configuration is adopted in the two plates as a compromise between the two initial configurations.
Different simulations have been carried out as a function of the perpendicular external magnetic field $H_{ext}$, starting from different initial configurations. The square and triangle curves of Figure 4 
(b) show the behavior of the pull-off force starting from stripe patterns of $0.09$ and $0.18\,\mu$m width respectively. At $H_{ext}=0$ we notice that larger stripes lead to a smaller pull-off force.
In terms of the effective model, we deal with the same situation as that of Figure 2 (f) but now at vanishing separation. This means that since the attractive 1--3 interaction dominates 
largely over any change of the repulsive 1--2 and 1--4 interactions caused by $H_{ext}$, the pull-off force is determined solely by the number of domain walls, i.e. by the total domain perimeter.
Until the stripes break down, no change in the total perimeter occurs, so there is a constant friction force until its sudden drop.
To verify this hypothesis, we performed a set of pull-off simulations starting from a saturated sample, at high negative $H_{ext}$, and decreasing the field leaving the domains free to nucleate and
expand until they reach a steady state with disordered maze-like patterns. The pull-off force at different $H_{ext}$ is reported in Figure 4 (b) (circles). Figure 4 (c) shows the 
corresponding total perimeter of the domains in the FFs and Figure 4 (d) displays the equilibrium magnetization in the lower film at slightly negative, at zero, and at positive values of $H_{ext}$. 
The sudden jump at negative fields is due to the nucleation of the domains, but at positive fields the decrease in the pull-off force is smoother than in the other two cases. 
The domain annihilation is in fact slower and more gentle with maze-like patterns than with stripe domains.
A linear proportionality between the pull-off force and the domain perimeter is found, and the maximum adhesion occurs for $H_{ext}=0$, where the demagnetization process requires the maximization of the domain perimeter.
As was done for the friction force, we can think of controlling the adhesive properties of two plates by acting on them with an external field: in movie 6 in Supporting Information, a force $W<W_{po}$ is applied to the upper plate for $H_{ext}=0$.
When $H_{ext}$ is turned on, the domains change their perimeter thus decreasing $W_{po}$ and causing the plates to detach.

\section*{Conclusions}
In this paper we proposed a model sliding system that, based on the interaction between magnetic domains, offers the possibility of a motion control without wear and degradation. This model can be exploited both to investigate the basic physics of friction and to design new mechanisms for sliding and adhesion control in micro-mechanical systems, avoiding the drawbacks arising from mechanical contact.\\ 
Simulating for the first time the dynamical properties of two interacting distributions of magnetic domains driven out of equilibrium, we have extensively discussed the 
role played by the geometric parameters and material properties in determining the mechanical response of the driven system.
We have also proved that in presence of domain plasticity only the simultaneous solution of both the upper plate's motion and the domain dynamics leads to a complete 
characterization of all possible sliding regimes.\\
In certain conditions, however, there is no need of considering such full dynamical interplay. For example, 
arbitrary domain patterns could be written on ferromagnetic films using techniques and materials' systems from the magnetic recording industry.  
These magnetic structures would be stable in our working conditions, i.e.~unaffected (``frozen-in'') by the magnetic fields applied and by thermal fluctuations.
One can thus 
easily control the domain geometry and periodicity, addressing problems related to structural lubricity and commensurability, and paving the way to controlling microscale motion solely exploiting magnetic friction.\\ 
 The calculated orders of magnitude for the distances and forces lie in the range detectable with an SFA \cite{meyer,carpick} or a meso-scale friction tester \cite{wang}. For small forces also 
an AFM could be used if equipped with a colloidal probe tip having a large radius of curvature and a soft cantilever. To measure the friction force on a genuine 
plane-on-plane geometry, an AFM can be equipped and operated as described by Tang et al. \cite{tang}. Measurements could be performed both in non-contact and contact mode. In the 
latter case, a non-magnetic spacing layer could be used to keep the two coating films at the proper distance with sub-nanometric precision. In contact mode, both the magnetic 
and mechanical friction could be simultaneously measured. In order for the former to prevail, the mechanical component could be lowered using a lubricant or by properly choosing a low friction 
material for the spacing layer. Depending on their size, the domains could be imaged prior and after the sliding with optical techniques as well as with magnetic force 
microscopy \cite{hubert}.\\
Further developments of the ideas illustrated in this work require an interdisciplinary effort, combining concepts and knowledge from magnetism and tribology and competences from data storage and micro-machining technologies.

\section*{Computational Method}
A detailed derivation of the theory hereafter presented can be found in \cite{pt}.
In our FFs with perpendicular anisotropy, the magnetization is assumed to be uniform along the $z$-axis.
In the approximation of thin domain walls, only its $z$ component plays a
relevant role, thus the domain dynamics can be described by a function varying only along the film plane, i.e. $\mathbf{m}_L\equiv M_s m_L(x,y)\hat{\mathbf{z}}$ and $\mathbf{m}_U\equiv M_s 
m_U(x,y)\hat{\mathbf{z}}$ for the lower and upper film respectively, $M_s$ is the saturation magnetization, $m_L$ and $m_U$ are dimensionless \cite{jagla1,jagla2}. 
This approximation allows a realistic description of the FFs and has proved to be successful in reproducing quantitatively many experimental 
features at both the macroscopic and microscopic level \cite{benassimag1,benassimag2}. With our assumptions, the system Hamiltonian for the isolated lower film can be written as: 
\begin{align}
\nonumber
\mathcal{H}_L&=\int d^3\mathbf{r}\Bigg\{ -K_u(\mathbf{r})\frac{m_L^2}{2}+\frac{A}{2}(\nabla_{\mathbf{r}}m_L)^2+
\frac{\mu_0 M_s^2}{4 \pi t_L }\\
\nonumber
&\times 
\int d^2\mathbf{r}'\Bigg[\frac{m_L(\mathbf{r}) m_L(\mathbf{r}')}{\vert\mathbf{r}-\mathbf{r}'\vert}-\frac{m_L(\mathbf{r}) m_L(\mathbf{r'})}{\sqrt{(\mathbf{r}-\mathbf{r}')^2+t_L^2}}
\Bigg]\\
 &- \mu_0 M_s m_L H_{ext} \Bigg\}
\label{hamilt}
\end{align}
with $d^3\mathbf{r}$ running over the FF volume and $d^2\mathbf{r}'$ spanning the $xy$-plane, the local magnetic field entering the LLGE can be obtained as $\mathbf{B}=-1/M_s\, \delta 
\mathcal{H[\mathbf{m}]}/\delta \mathbf{m}$.
The first term in Equation (\ref{hamilt}) is the anisotropy energy, with $K_u(\mathbf{r})=K_u \vert 1-\eta P(\mathbf{r}) \vert$ 
where $K_u$ is the macroscopic measurable uniaxial anisotropy constant and $P$ is a unitary Gaussian white noise term producing anisotropy fluctuations of strength $\eta$ throughout 
the sample. These fluctuations represent the film inhomogeneities and defects which pin the domains.
The second term is the exchange energy, with $A$ (exchange stiffness constant) representing the energy cost of a local variation of the magnetization. 
The third term represents the long-range, non-local, stray field with $\mu_0$ the vacuum permeability and $M_s$ the saturation magnetization of the film: this term is responsible for the spontaneous demagnetization of the film.
The last term is the energy due to the presence of an external magnetic field perpendicular 
to the film plane.
A similar Hamiltonian can be written for the upper film.\\
When the two films are brought closer, each one feels the magnetic field due to the other and the following interaction energy arises:
\begin{align}
\nonumber
&\mathcal{H}_{INT}=\frac{\mu_0 M_s^2}{4 \pi}\int d^2\mathbf{r}\int d^2\mathbf{r}'\bigg(
-\frac{m_U(\mathbf{r}')m_L(\mathbf{r})}{\sqrt{(\mathbf{r}-\mathbf{r}')^2+d^2}}+\\
\nonumber
&\frac{m_U(\mathbf{r}')m_L(\mathbf{r})}{\sqrt{(\mathbf{r}-\mathbf{r}')^2+(d+d_U)^2}}+
\frac{m_U(\mathbf{r}')m_L(\mathbf{r})}{\sqrt{(\mathbf{r}-\mathbf{r}')^2+(d+d_L)^2}}-\\
&\frac{m_U(\mathbf{r}')m_L(\mathbf{r})}{\sqrt{(\mathbf{r}-\mathbf{r}')^2+(d+d_L+d_U)^2}}
\bigg).
\label{inter}
\end{align}
Notice the simultaneous dependence on $m_L$, $m_U$ and $d$.
The simulated domain behavior as a function of the separation $d$ is in good agreement with the available data and with the analytical results \cite{baltz}.\\
So far we have only modelled the domain evolution inside the two interacting FFs. To simulate the sliding motion of the upper film, we have to calculate the force that the lower film exerts on it.
The force that a magnetic field exerts on a magnetic dipole moment can be defined in two ways depending on the nature of the dipole itself \cite{boyer}, however our magnetic field $\mathbf{B}$ is irrotational and the two definitions coincide leaving no ambiguities.
The total force acting on the upper film can be thus calculated as the sum over all the infinitesimal forces acting on each of its dipole moments:
\begin{align}
F_i=\int d^3\mathbf{r}\; m_U(\mathbf{r})\frac{\partial B_i(\mathbf{r},m_L)}{\partial z}.
\label{inter}
\end{align}
Here, too, the force depends simultaneously on $m_L$, $m_U$ and $d$.\\
A well known problem with molecular dynamics simulations of atomic-scale friction it that the choice of $v_0$ is limited by the need of sampling both the fast phonon dynamics and the slower stick-slip sliding in a reasonable simulation time \cite{vanossi}. 
It is also well known that a satisfactory description of the tribological properties can be achieved with a choice of slider mass $m$ and driving velocity $v_0$ that decouples 
the fast atomic motion from the slower slider dynamics, even if the resulting $m$ and $v_0$ values are orders of magnitude far from the experimental ones. In our magnetic counterpart, the magnetization dynamics replaces the atomic motion in being 
much faster than stick-slip sliding. With the choice $v_0=5$ m/s, $m=1.6\times 10^{-19}$ Kg and $k=0.15$ N/m, we can reasonably decouple the magnetization and the slider dynamics in the stick-slip regime.
For the time integration of the LLGEs we used the semi-implicit first order algorithm described in \cite{pt}. To circumvent the problem of the non-locality of the Hamiltonian, the LLGE have been treated in reciprocal space assuming periodic 
boundary conditions for the magnetization. The slider equation of motion has been integrated with a simple Velocity-Verlet algorithm also assuming periodic boundary conditions in the $xy$-plane.\\
Magnetic domains have no inertia, thus the LLGE is by definition an overdamped equation \cite{gilbert,brown}. The damping represents the energy dissipation occurring through microscopic degrees of freedom such as phonons, magnons and eddy currents in
conducting FFs. The damping coefficient is the inverse of the characteristic microscopic relaxation time of the system. In our case we have chosen $\tau=10^{-9}$ s, in reasonable agreement with recent experiments \cite{mizukami}.
Also the slider equation of motion takes into account the energy dissipated through the driving apparatus by means of a viscous damping term $- m \gamma \mathbf{v}(t)$ where $\mathbf{v}(t)$ is the upper plate velocity. The choice of $\gamma$ is again dictated 
by the necessity of sampling many stick-slip events in a reasonable simulation time and is related to the choice of $m$ and $v_0$. Choosing $1/\gamma=\tau$ we can damp the mechanical energy of the slider in a time larger than the slip time and smaller than the stick time, as
occurs in a realistic situation.\\
In demagnetized films such as Co/Pt multilayers the stability of the domains is not significantly affected by thermal fluctuations. This follows from the lack of strong sample stray fields in the demagnetized state, the strong anisotropy, and the fact that the film’s grains are not 
decoupled. Indeed, no spontaneous domain wall motion at room temperature and zero applied fields are observed for these films with MFM down to $10$ nm scales. Accordingly, we are justified in carrying out the simulations at $T=0$ K.\\
While in the middle of a domain the infinitesimal dipole moments are completely aligned along the $z$ direction, perpendicular to the FF surface, inside a domain wall they rotate, having a significant   
component within the $xy$-plane. The domain wall regions are thus sensitive to the presence of a magnetic field parallel to the FF surface, and the latter can be used to order the maze-like patterns into parallel stripes.
Our model focusses only on the $z$ component of the magnetization, thus we cannot simulate the domain ordering process over time. However, we can start the simulations with the domains already ordered into stripes of a given width (i.e. a given stripe density) and let them evolve
in time to see whether they remain stable or not. If we are too far from the stripe width that minimizes the energy, the stripes will immediately break into pieces or merge together.
We can thus set up a trial and error procedure to obtain realistic stripe patterns to be used as an
initial configuration for the sliding simulations. Our procedure is not capable of identifying a unique optimal $\ell$ minimizing the total energy, rather we have a window of $\ell$ values for which the stripe pattern is stable. This happens also in real FFs where, depending on the 
alignment procedure, different widths are found for stable stripe patterns. This freedom of choice for the stripe width, both in the modeling and in the experiments, comes from the energy landscape of this kind of systems which is glassy, due to the pinning 
inhomogeneities of the FFs, with a very shallow energy minimum as a function of $\ell$.
 
\section*{Acknowledgements}
This work has been supported by grant CRSII2 136287/1 from the Swiss National Science Foundation.

\end{document}